\documentstyle[11pt,moriond,epsfig,amssymb]{article}

\def\Journal#1#2#3#4{{#1} {\bf #2} (#4) #3}
\def\NIM{{\em Nucl. Instr. and Meth.} A}
\def\NPB{{\em Nucl. Phys.} B}

\def\PRD{{\em Phys. Rev.} D}
\def\PRX{{\em Phys. Rev.}}
\def\ZPC{{\em Z. Phys.} C}
\def\ZPX{{\em Z. Phys.}}
\def\EPC{{\em Eur. Phys. Jour.} C}
\def\CPC{{\em Comp. Phys. Comm.}}

\def\gp{\gamma p}
\def\q2{Q^2}
\def\kt{k_T}
\def\ica{iterative cone algorithm}
\def\kca{$\kt$ cluster algorithm}
\def\pb1{pb$^{-1}$}
\def\etjet{E_T^{jet}}
\def\etajet{\eta^{jet}}
\def\yr{0.2<y<0.85}
\def\qr{\q2\leq 4}
\def\qrr{\q2\leq 1}
\def\g2{GeV$^2$}
\def\set{d\sigma/d\etjet}
\def\etjb{E^{jet}_{T,{\rm Breit}}}
\def\etajb{\eta^{jet}_{\rm Breit}}
\def\etalab{\eta^{jet}_{\rm LAB}}
\def\mj{M^{JJ}}
\def\cost{\vert\cos\theta^*\vert}
\def\smj{d\sigma/d\mj}
\def\scost{d\sigma/d\cost}

\begin{document}
\pagestyle{plain}
\title{JET PHYSICS AT HERA$^\dag$}
\footnotetext{$^\dag$Talk given at the {\it XXXIVth Rencontres de Moriond},
QCD and High Energy Hadronic Interactions, Les Arcs, Savoie, France,
March $20^{th}-27^{th}$, 1999.}

\author{C. GLASMAN, for the H1 and ZEUS Collaborations}

\address{Departamento de F\'\i sica Te\'orica C-XI,
Facultad de Ciencias,
Universidad Aut\'onoma de Madrid,\\
Cantoblanco, 28049 Madrid, Spain}

\maketitle\abstracts{
Measurements of inclusive jet and dijet cross sections in photoproduction
and deep inelastic scattering are presented. These measurements provide new
tests of QCD, constrain the parton densities of the proton and the photon, and
allow the search for new physics. Measurements of jet shapes are reported and
used to test the differences between quark and gluon jets.}

\section{\bf Introduction}
The main sources of jets at HERA are deep inelastic scattering (DIS) and 
quasi-real photon-proton ($\gp$) collisions (photoproduction). In DIS, a
highly virtual photon ($\q2\gg 0$, where $\q2$ is the virtuality of the
exchanged photon) interacts with a parton from the proton. In hard
photoproduction, a parton from the proton interacts with a quasi-real photon
($\q2\sim 0$) emitted by the electron beam. High-$p_T$ jet production is
sensitive to the parton densities in the proton, in particular the gluon
density, and to the parton densities in the photon in $\gp$ interactions. The
measurements of jet cross sections provide constraints on the proton and
photon parton densities, constitute new tests of QCD and allow the search for
new physics. The internal structure of a jet is another tool to investigate
QCD predictions and is sensitive to the differences between quark and gluon
jets. During $1994-1997$ HERA operated with positrons of energy
$E_e=27.5$~GeV colliding with protons of energy $E_p=820$~GeV.

\section{\bf Photoproduction at HERA}
In $\gp$ interactions, the cross section for the production of jets at leading
order (LO) in perturbative QCD (pQCD) is given by two contributions, namely
those of resolved and direct processes. In resolved processes, a parton
from the photon interacts with a parton from the proton, whereas in direct
processes, the photon interacts as a point-like particle with a parton from the
proton. The two contributions to the jet production cross section can be
written as

\begin{equation}
\sigma_{direct}=\int d\Omega\ f_{\gamma /e}(y)\ f_{j/p}(x_p,\mu^2_F)\
d\sigma(\gamma j\rightarrow {\rm jet}\ {\rm jet})
\end{equation}

and

\begin{equation}
\sigma_{resolved}=\int d\Omega\ f_{\gamma /e}(y)\ f_{i/\gamma}(x_{\gamma},\mu^2_F)\
f_{j/p}(x_p,\mu^2_F)\
d\sigma(ij\rightarrow {\rm jet}\ {\rm jet}),
\end{equation}

where $f_{\gamma /e}(y)$ is the flux of photons in the
electron \footnote{The variable $y$ is the fraction of the electron
energy taken by the photon.} which is usually estimated by the
Weizs\"acker-Williams approximation~\cite{wwa}; $f_{j/p}(x_p,\mu^2_F)$ are the
parton densities in the proton (determined from e.g. global fits~\cite{mrs}),
$x_p$ is the fraction of the proton momentum taken by the parton and
$\mu_F$ is the factorisation scale; and
$d\sigma(\gamma(i)j\rightarrow {\rm jet}\ {\rm jet})$ is the subprocess cross
section, calculable in pQCD. In the case of resolved processes,
there is an additional ingredient: $f_{i/\gamma}(x_{\gamma},\mu^2_F)$ are the
parton densities in the photon, for which there is only partial information.
The integral is performed over the phase space represented by ``$d\Omega$''.

\section{\bf Inclusive jet cross sections in photoproduction}
Inclusive jet cross sections in $\gp$ interactions are sensitive to the
underlying parton dynamics and to the parton densities in the photon. To test
the available parametrisations of the photon parton densities, the
experimental and theoretical uncertainties must be reduced as much as
possible. Two approaches have been followed to reduce the experimental
uncertainties due to the so-called ``underlying event''~\cite{highet,nh1}:
e.g. to decrease the cone radius in iterative cone jet
algorithms~\cite{cone,snow} and/or to increase the jet transverse
energy~\cite{highet}. In this way, a region of phase space was found where the
next-to-leading order (NLO) QCD calculations are able to describe the
data~\cite{highet}. However, the uncertainties in the calculations for the
\ica\ are still sizeable. The use of the $\kt$ cluster algorithm~\cite{kt}
reduces the theoretical uncertainties since it allows a transparent
translation of the theoretical jet algorithm to the experimental set-up by
avoiding the ambiguities related to the merging and overlapping of jets. In
addition, a meaningful comparison between data and NNLO QCD calculations (when
available) for the \kca\ will be possible.

Inclusive jet cross sections have been measured~\cite{incjet} using the
$1995-1997$ ZEUS~\cite{status} data (which amounts to an integrated
luminosity of ${\cal L}\sim 40$ \pb1) as a function of
the jet transverse energy ($\etjet$). The jets have been searched for with the
\kca. The measurements have been performed for jets of hadrons with $\etjet$
between 17 and 74 GeV and jet pseudorapidity ($\etajet$) between $-0.75$ and
$2.5$, and are given for the kinematic region defined by $\yr$ and $\qr$ \g2.

\begin{figure}[t]
\vspace{-2cm}
\centerline{\mbox{
\epsfig{figure=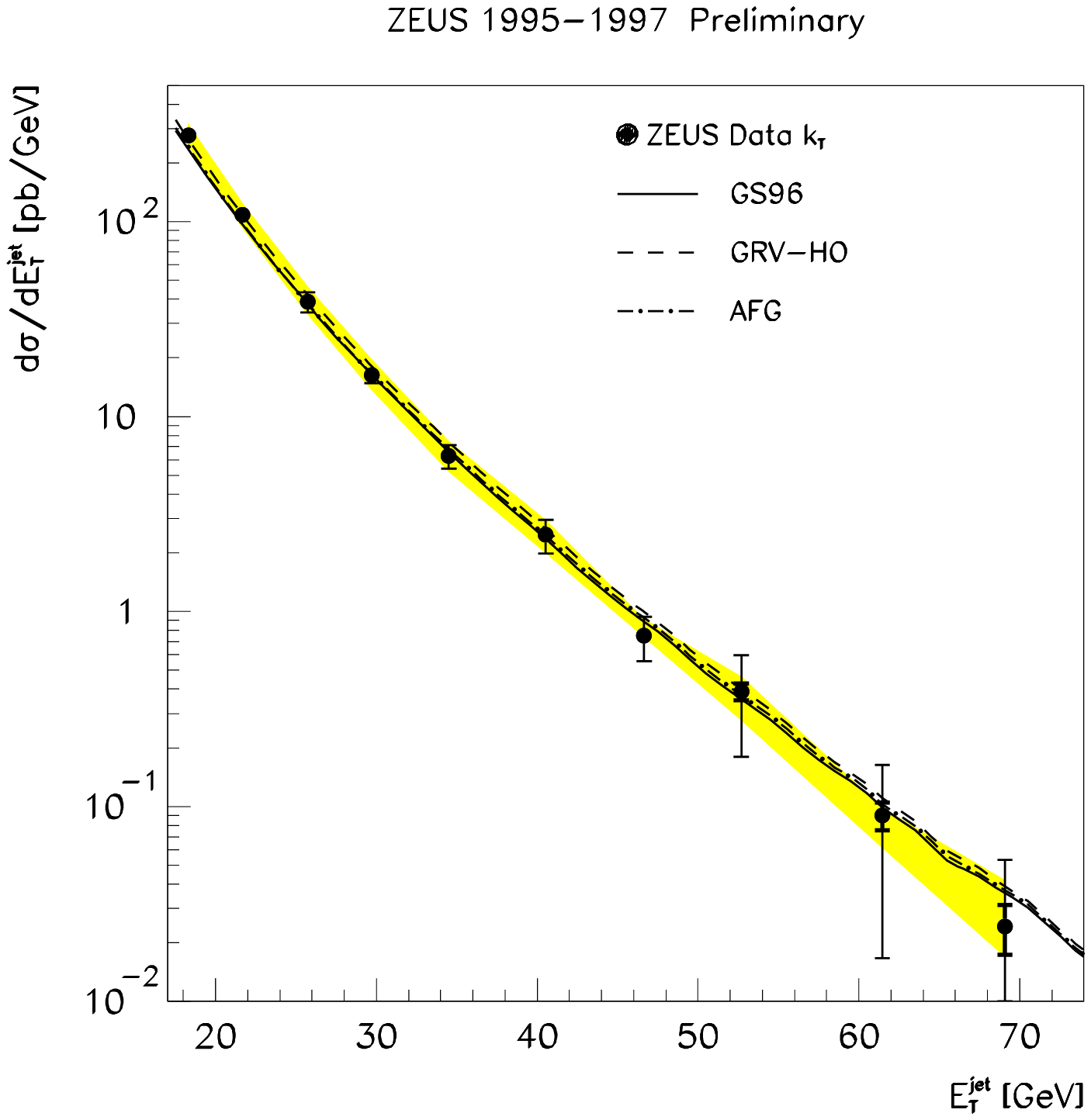,width=10cm}
\hspace{-2cm}
\epsfig{figure=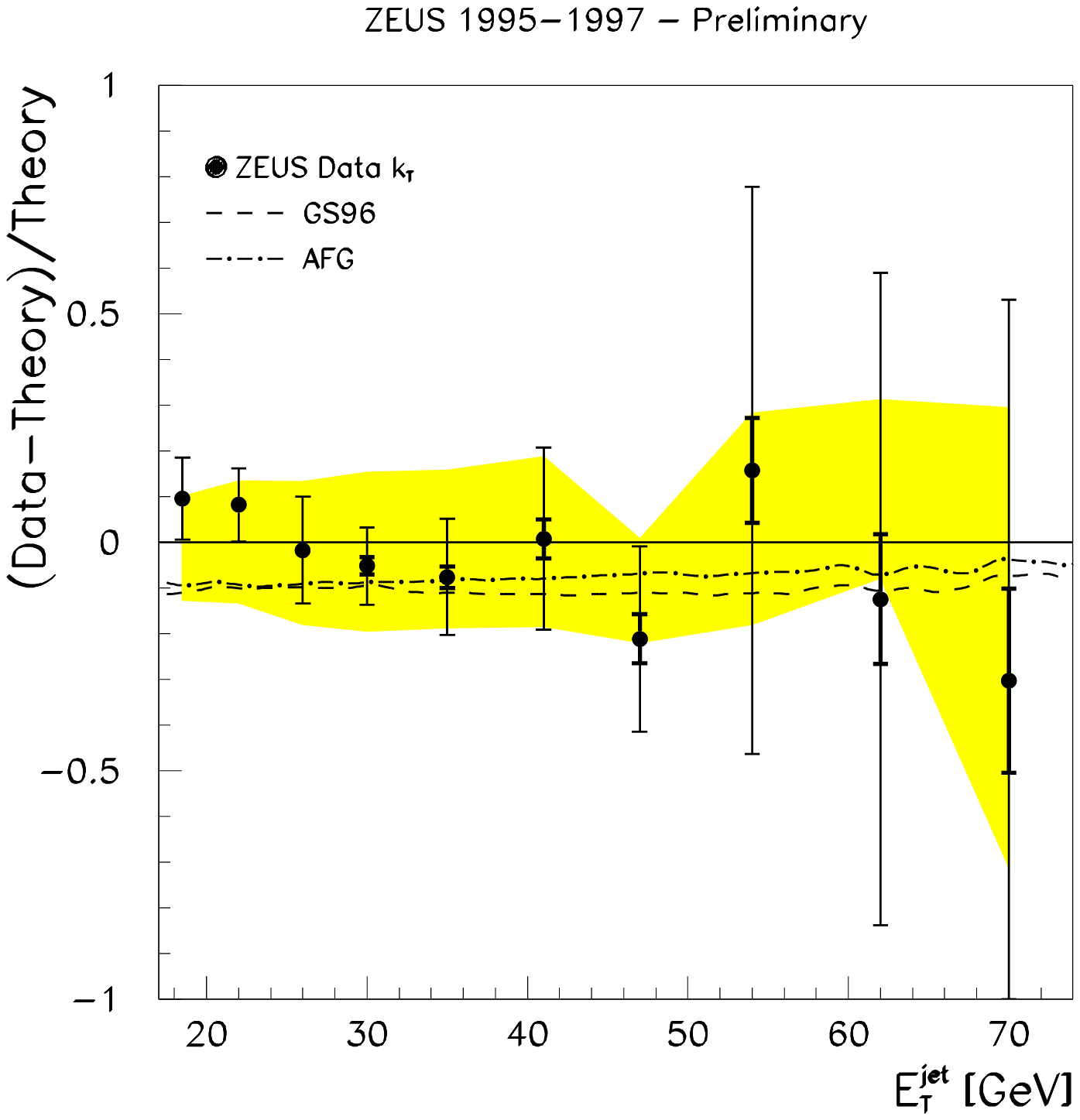,width=10cm}}}
\vspace{-1cm}
\caption{\label{one}{Measurement of the inclusive jet cross section $\set$ in
photoproduction. NLO QCD calculations are shown for comparison.}}
\end{figure}

Figure \ref{one} shows the measured $\set$ (black dots). In figures \ref{one}
and \ref{two}, the systematic uncertainties not associated with the absolute
energy scale of the jets have been added in quadrature to the statistical
errors (thick error bars) and are shown as thin error bars. The shaded band
represents the uncertainty on the energy scale of the jets. The data show a
steep fall-off of four orders of magnitude in the measured range. The curves
are NLO QCD calculations~\cite{klasen,harris} using different parametrisations
of the photon structure function: GS96~\cite{gs} (solid line),
GRV-HO~\cite{grv} (dashed line) and AFG~\cite{afg} (dot-dashed line). The
CTEQ4M~\cite{cteq4} proton parton densities have been used in all cases.
In the calculations shown here, the renormalisation and factorisation
scales have been chosen equal to $\etjet$ and $\alpha_s$ was calculated
at 2-loops with $\Lambda^{(4)}_{\overline{MS}}=296$ MeV. The NLO
calculations give a reasonable description of the data. The right-hand side
of figure \ref{one} shows the fractional differences \footnote{The
fractional differences are taken with respect to the NLO calculations
based on GRV-HO.} between the measured $\set$ and the NLO calculations.
The NLO QCD calculations using the current knowledge of the photon
structure are able to describe the data within the present experimental and
theoretical uncertainties.

\section{\bf High-mass dijet cross sections in photoproduction}
The dijet mass distribution $\mj$ is sensitive to the presence of new
particles or resonances that decay into two jets. The distribution of the
angle between the jet-jet axis and the beam direction in the dijet
centre-of-mass system ($\cos\theta^*$) reflects the underlying parton dynamics
and is sensitive to the spin of the exchanged particle. New particles or
resonances decaying into two jets may also be identified by deviations in the
measured $\cost$ distribution with respect to the QCD predictions.

High-mass dijet cross sections have been measured~\cite{dijet} using the
$1995-1997$ ZEUS data as a function of $\mj$ and $\cost$ in the
kinematic region given by $\yr$ and $\qr$ \g2. The measurements have been
performed for $\mj>47$ GeV and $\cost<0.8$ using the \kca.

\begin{figure}[t]
\vspace{-2cm}
\centerline{\mbox{
\epsfig{figure=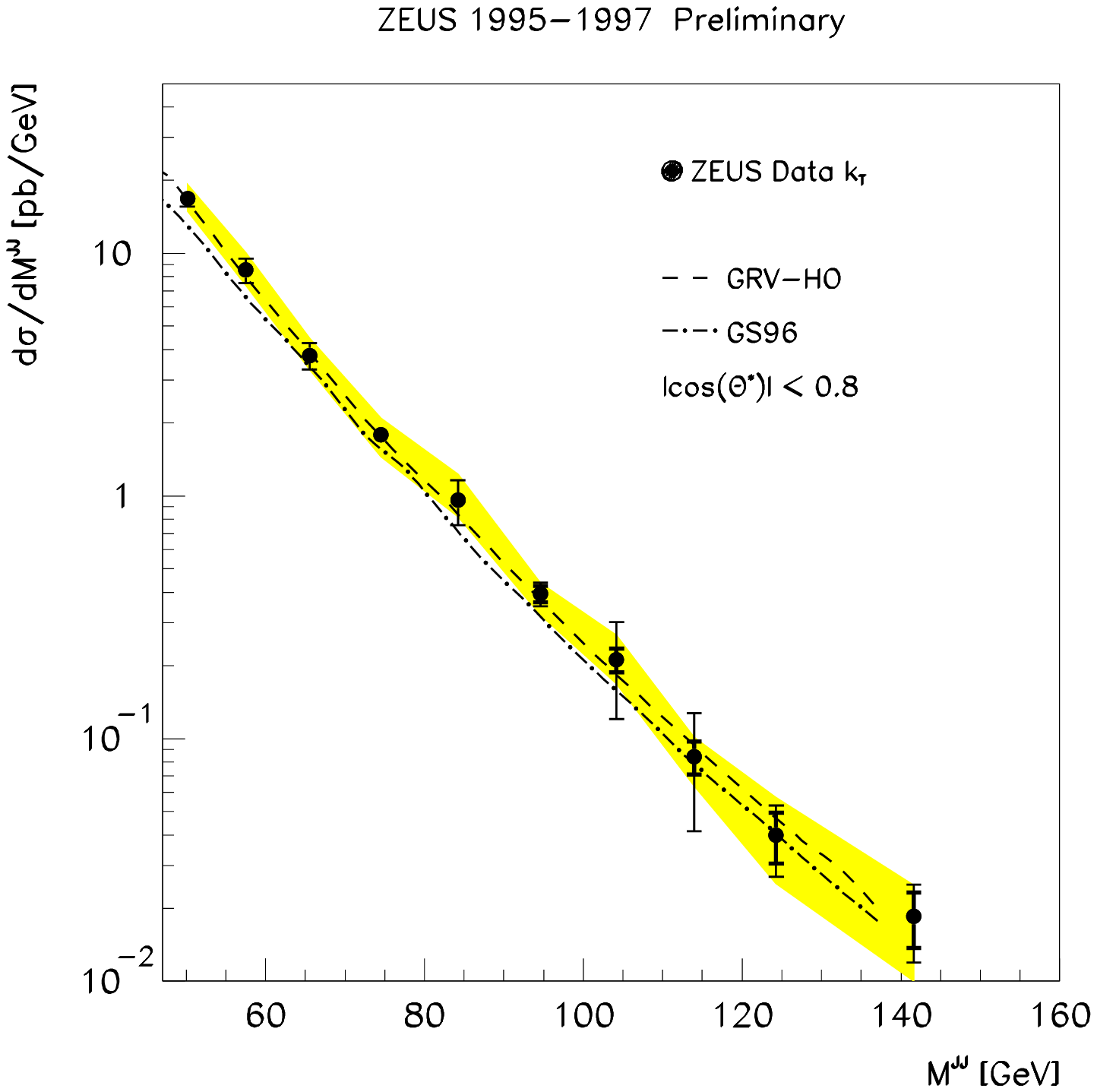,width=10cm}
\hspace{-2cm}
\epsfig{figure=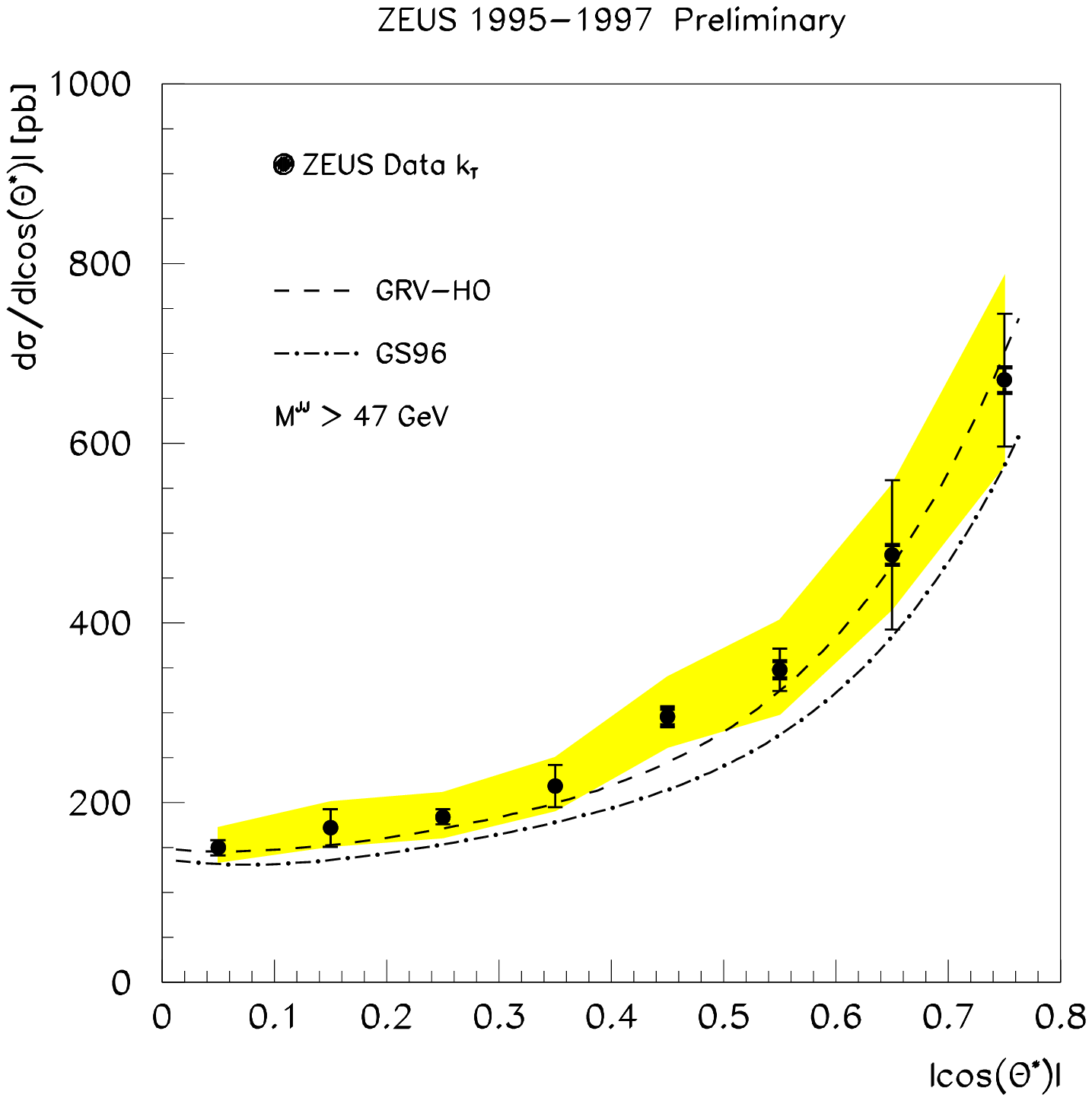,width=10cm}}}
\vspace{-1cm}
\caption{\label{two}{Measurement of the high-mass dijet cross sections $\smj$
and $\scost$ in photoproduction. NLO QCD calculations are shown for
comparison.}}
\end{figure}

Figure \ref{two} shows the measured $\smj$ (left-hand side) and $\scost$
(right-hand side). The data show a steep fall-off in $\mj$ of three orders
of magnitude in the measured range. The measured $\scost$ rises as $\cost$
increases. The NLO QCD calculations~\cite{klasen} give a reasonable
description of the measured distributions. The predictions based on GRV-HO are
closer in magnitude to the measured cross sections. No significant deviation
between data and NLO calculations is observed in the measured range of $\mj$
and $\cost$.

\section{\bf Jet shapes in photoproduction}
The internal structure of the jets has been studied by means of the
differential and the integrated jet shape. The differential jet shape
($\rho(r)$) is defined as the averaged fraction of the jet transverse energy
which lies in the annulus of width $\Delta r$ at a distance $r$ from the jet
axis,

\begin{equation}
\rho(r)=\frac{1}{N_{jets}\Delta r}
\displaystyle\sum_{jets}
\frac{E_T(r-{\Delta r\over 2},r+{\Delta r\over 2})}{E_T(0,1)},
\end{equation}

where $E_T(r-\Delta r/2,r+\Delta r/2)$ is the transverse energy within the 
given annulus and $N_{jets}$ is the total number of jets in the sample.
The integrated jet shape ($\psi(r)$) is defined as

\begin{equation}
\psi(r)=\frac{1}{N_{jets}}\displaystyle\sum_{jets}
\frac{E_T(0,r)}{E_T(0,1)}.
\end{equation}

QCD predicts that at sufficiently high transverse energy, the jet
shape is driven by gluon emission off the primary parton and therefore
calculable in pQCD. Gluon jets are predicted to be broader than quark jets due
to the gluon-gluon coupling strength being larger than that of the quark-gluon.
At LO, a jet consists of only one parton and thus has no structure. NLO QCD
calculations give the lowest non-trivial order contribution to the jet
structure. 

The differential and integrated jet shapes have been measured for jets defined
with the \kca\ and required to have $\etjet>17$ GeV and $-1<\etajet<2$. The
measurements have been performed using the 1995 ZEUS data (${\cal L}\sim 6$
\pb1) in the kinematic region given by $\yr$ and $\qrr$ \g2.

\begin{figure}[t]
\vspace{-2cm}
\centerline{\mbox{
\hspace{2.5cm}
\epsfig{figure=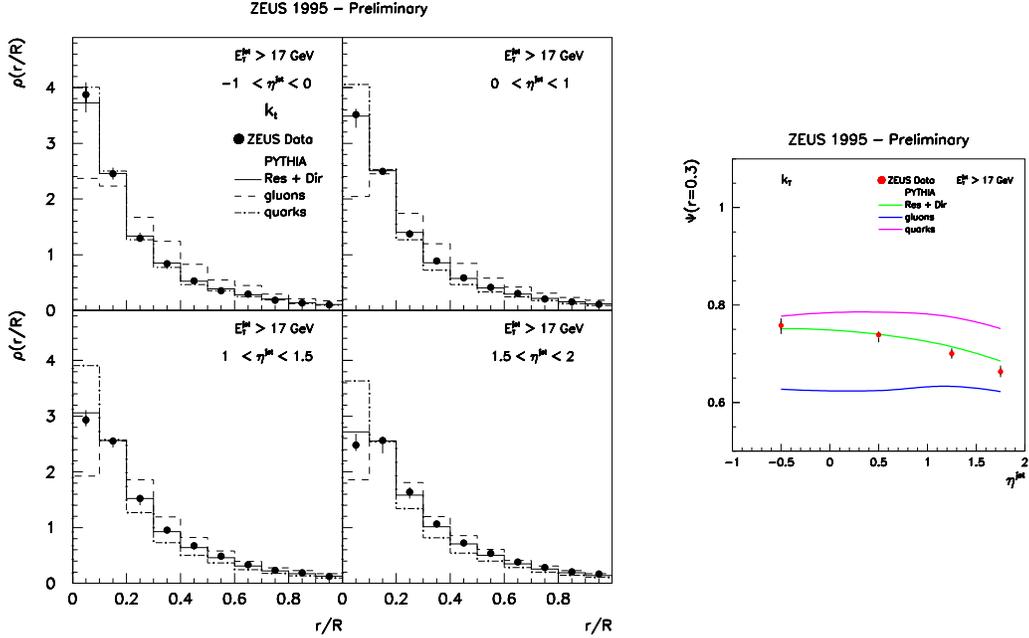,width=20cm}}}
\vspace{-16cm}
\caption{\label{three}{Measured differential (left-hand side) and integrated
(right-hand side) jet shapes in photoproduction. Calculations from the
leading-logarithm parton-shower Monte Carlo PYTHIA are shown for comparison.}}
\end{figure}

Figure \ref{three} shows the measured differential jet shape $\rho(r)$ as a
function of $r$ in different regions of $\etajet$ (left-hand side). The
data (black dots) show that the jets become broader as $\etajet$ increases.
The solid histograms are the predictions of a leading-logarithm
parton-shower Monte Carlo (MC) calculation using PYTHIA~\cite{pythia} for
resolved and direct processes. The calculations, which include initial- and
final-state QCD radiation, give a good description of the data. The dashed
(dot-dashed) histograms are the predictions of PYTHIA for samples of gluon
(quark) jets. The measured jets are dominated by quarks for $-1<\etajet<0$, and
become increasingly more gluon-like as $\etajet$ increases.

The quark and gluon content of the final-state jets has been studied in more
detail by looking at the integrated jet shape for a fixed value of $r$
($r=0.3$) as a function of $\etajet$ (right-hand side of figure \ref{three}).
The measured jet shape decreases with $\etajet$, i.e. the jets become broader
as $\etajet$ increases. Comparing the data to the model predictions for gluons
(lower curve) and quarks (upper curve), it is observed that the data go from
being dominated by quarks to being dominated by gluons as $\etajet$ increases.
Thus, the broadening of the jets is consistent with an increasing fraction of
gluon-initiated jets as $\etajet$ increases.

\section{\bf Deep inelastic scattering at HERA}
Two processes contribute to the dijet cross section in neutral-current DIS at
LO in pQCD, namely the QCD Compton ($\gamma^*q\rightarrow qg$) and the Boson
Gluon Fusion (BGF) ($\gamma^* g\rightarrow q\bar q$) processes. The cross
section for jet production is given by

\begin{equation}
\sigma_{DIS}=\int d\Omega\ f_{j/p}(x_p,\mu_F^2)\ 
d\sigma_{\gamma^*j\rightarrow {\rm jet}\ {\rm jet}}(x_p\cdot P,\alpha_s(\mu_R^2),\mu_R^2,\mu_F^2),
\end{equation}

where $f_{j/p}(x_p,\mu_F^2)$ are the parton densities in the proton and
$d\sigma_{\gamma^*j\rightarrow {\rm jet}\ {\rm jet}}$ is the subprocess cross
section. In the kinematic regimem studied, the BGF process represents the
largest contribution and therefore the final state is expected to be dominated
by quark-initiated jets.

\section{\bf Jet shapes in DIS}
The integrated jet shape has been measured~\cite{disshape} using the
1994 H1~\cite{statush1} data (${\cal L}\sim 2$ \pb1) for dijet events
in the Breit frame. The jets have been searched for with the iterative cone
and the $\kt$ cluster algorithms. The two jets with highest $\etjb$
(transverse with respect to the direction of the virtual photon in the Breit
frame) which satisfy $\etjb>5$ GeV and $-1<\etalab<2$ have been selected. The
measurements are presented for different regions of $\etajb$ and $\etjb$ in
the kinematic region given by $10<\q2\lesssim 120$ \g2\ and
$2\cdot 10^{-4}\lesssim x_{Bj}\lesssim 8\cdot 10^{-3}$, where $x_{Bj}$ is the
Bjorken $x$ variable.

\begin{figure}[t]
\vspace{-6cm}
\centerline{\mbox{
\hspace{2.5cm}
\epsfig{figure=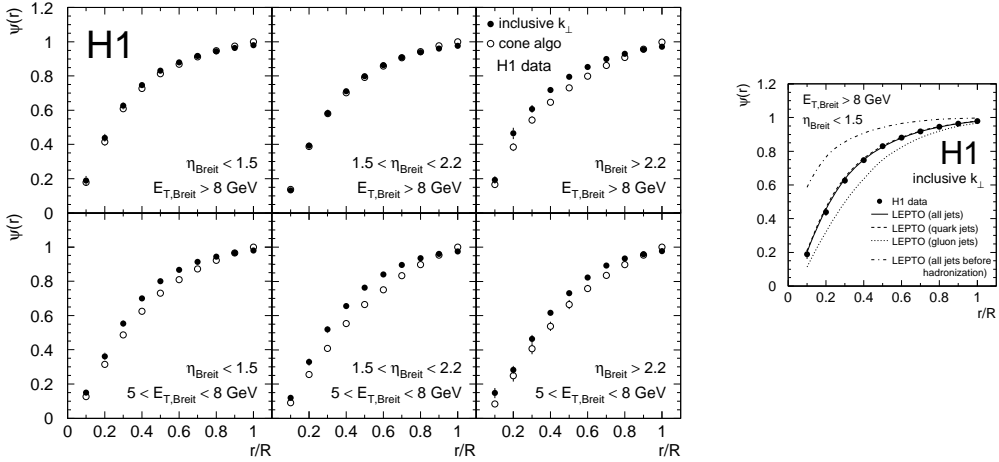,width=20cm}}}
\vspace{-13.5cm}
\caption{\label{four}{Measured integrated jet shapes in DIS. On the right-hand
side, calculations from the leading-logarithm parton-shower Monte Carlo LEPTO
are shown for comparison.}}
\end{figure}

Figure \ref{four} shows the measured integrated jet shape as a function of $r$
in different regions of $\etajb$ and two regions of $\etjb$ for the two
jet algorithms. The data show that the jets get narrower as $\etjb$
increases and that the jets get broader as $\etajb$ increases (i.e.
towards the proton direction). The broadening of the jets is more evident
at low $\etjb$. At low $\etjb$, the jets defined by the \kca\ are
narrower than those defined by the \ica. At high $\etjb$ both
algorithms produce very similar jets.

The QCD based MC LEPTO~\cite{lepto} predicts that gluon jets (dotted
line in the right-hand side of figure \ref{four}) are broader than quark jets
(dashed line) and that in the kinematic region of the measurements, the dijet
sample is dominated by BGF and therefore by $q\bar q$ pairs in the final
state. The comparison of the model predictions to the data shows that the
measured jet shape is compatible with that of quark-initiated jets.

\section{\bf Dijet cross sections in DIS}
Dijet cross sections in neutral-current DIS have been measured~\cite{discross}
using the $1994-1997$ H1 data (${\cal L}\sim 36$ \pb1). The jets are defined by
the \kca\ in the Breit frame. The two jets with highest $\etjet$ are required
to have $\etjb>5$ GeV, $E^{jet1}_{T,{\rm Breit}}+E^{jet2}_{T,{\rm Breit}}>17$
GeV and $-1<\etalab<2.5$. The measurements of $d^2\sigma/d\q2 d\xi$, where
$\xi=x_{Bj}(1+M_{JJ}^2/\q2)$, and $d^2\sigma/d\q2 dx_{Bj}$ are presented for
the kinematic region given by $10<\q2<5000$ \g2\ and $0.2<y<0.6$. The data
have been corrected for detector effects, and for initial- and final-state QED
radiation effects.

For the inclusive \kca, the hadronisation effects are small and
independent of $\q2$. Thus, the data have been compared to the NLO
calculations performed using the program DISENT~\cite{disent}.

\begin{figure}[t]
\vspace{-5cm}
\centerline{\mbox{
\hspace{1cm}
\epsfig{figure=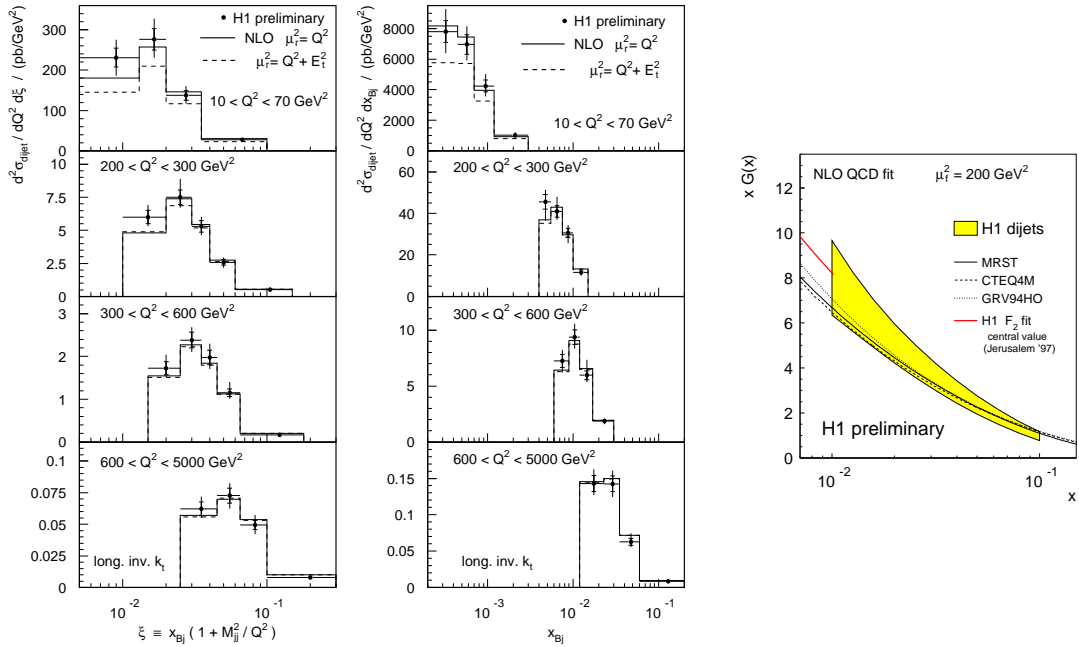,width=20cm}}}
\vspace{-12cm}
\caption{\label{five}{Measured dijet cross sections $d^2\sigma/d\q2 d\xi$ and
$d^2\sigma/d\q2 dx_{Bj}$ in DIS (left-hand side). NLO QCD calculations are shown
for comparison. On the right-hand side, the extracted gluon density in the
proton is shown.}}
\end{figure}

The left-hand side of figure \ref{five} shows the measured dijet cross
sections in DIS as a function of $\xi$ and $x_{Bj}$ in different regions of
$\q2$ (black dots). The histograms are the NLO calculations from DISENT for
two choices of the renormalisation scale: $\mu_R^2=\q2$ (solid histograms) and
$\mu_R^2=\q2+E_T^2$ (dashed histograms). The calculations based on different
choices of $\mu_R^2$ differ only at low $\q2$. The dependence of the
calculations on the factorisation scale is small. The NLO predictions
give a reasonable description of the data at all values of $\q2$ for the
choice $\mu_R^2=\q2$.

\section{\bf Extraction of the gluon density in the proton}
A direct determination~\cite{discross} of the gluon density has been
performed via a NLO QCD fit to the measured dijet cross sections. Inclusive
neutral-current DIS data~\cite{incdis} have been used to constrain the
quark densities. The determination of the gluon density has been done using
the dijet cross sections for $\q2>200$ \g2, where the hadronisation
corrections are small and independent of $\q2$, $\xi$ and $x_{Bj}$. The
theoretical uncertainties are also small in this kinematic region. The value
for the strong coupling constant has been taken from the world average,
$\alpha_s(M_Z^2)=0.119\pm 0.005$. The parton densities have been extracted
at a factorisation scale $\mu^2_F=200$ \g2\ in the range $0.01<x<0.1$.

The resulting quark densities are $5\%$ higher at $x=0.01$ than the results
from global fits and they are in agreement at $x=0.1$. The resulting gluon
density together with its uncertainties is shown as a shaded band in the
right-hand side of figure \ref{five}. This result is slightly higher than those
from the global analyses, although compatible within the errors. It is in good
agreement with the determination of the gluon density from the QCD
analysis~\cite{h1f2} of $F_2$ data at $x=0.01$ (dark solid line) by the H1
Collaboration.

\section{\bf Summary and conclusions}
Measurements of high-$E_T$ inclusive jet and dijet cross sections in
photoproduction have been presented.
NLO QCD calculations give a reasonable description of the measured
cross sections up to $\etjet\sim 74$ GeV. No significant deviation
between data and calculations is observed up to dijet masses of
$\mj\sim 140$ GeV.

The measurements of jet shapes in photoproduction and neutral-current DIS are
consistent with the different parton contents of the final state expected in
these two reactions: quark jets in DIS ($\gamma^*g\rightarrow q\bar q$ is the
dominant subprocess), and an increasing fraction of gluon jets as $\etajet$
increases in photoproduction (the resolved subprocess
$q_{\gamma}g_p\rightarrow qg$ dominates over that of the direct
$\gamma g\rightarrow q\bar q$ at large $\etajet$).

A direct determination of the gluon density in the proton has been
made via a QCD fit to the dijet and inclusive neutral-current DIS data.
The resulting gluon density is compatible within errors with the results from
global analyses, supporting the universality of the gluon density.

\section*{Acknowledgments}
I would like to thank the organisers for a very warm atmosphere and a
well organised conference. Special thanks to my colleagues from H1 and
ZEUS for their help in preparing this work. The work of the author is
supported by an European Community fellowship under contract ERBFMBICT
972523.

\end{document}